\documentclass[twocolumn]{aastex63}

\newcommand{\rion}{{$R_{\rm ion}$ }}
\newcommand{\simba}{{\sc Simba}}

\accepted{}
\usepackage[caption=false]{subfig}
\shorttitle{Reionization with {\sc Simba}}
\shortauthors{Hassan et al.}
\graphicspath{{./}{figures/}}
\begin{document}
\title{Reionization with {\sc Simba}: How much does astrophysics matter in modeling cosmic reionization?}

\correspondingauthor{Sultan Hassan}
\email{shassan@flatironinstitute.org}

\author[0000-0002-1050-7572]{Sultan Hassan}
\affiliation{Center for Computational Astrophysics, Flatiron Institute, 162 5th Ave, New York, NY 10010, USA}\affiliation{Department of Physics \& Astronomy, University of the Western Cape, Cape Town 7535,
South Africa}

\author{Romeel Dav\'e}
\affiliation{
Institute for Astronomy, University of Edinburgh, Royal Observatory, Edinburgh EH9 3HJ, UK}\affiliation{Department of Physics \& Astronomy, University of the Western Cape, Cape Town 7535,
South Africa}\affiliation{
South African Astronomical Observatories, Observatory, Cape Town 7925, South Africa}

\author{Matthew McQuinn}\affiliation{Department of Astronomy, University of Washington, Seattle, WA 98195}

\author{Rachel S. Somerville}\affiliation{Center for Computational Astrophysics, Flatiron Institute, 162 5th Ave, New York, NY 10010, USA}

\author{Laura C. Keating}\affiliation{Leibniz-Institute for Astrophysics Potsdam (AIP), An der Sternwarte 16,
14482 Potsdam, Germany}

\author{Daniel Angl\'es-Alc\'azar}\affiliation{Department of Physics, University of Connecticut, 196 Auditorium Road, U-3046, Storrs, CT, 06269, USA}\affiliation{Center for Computational Astrophysics, Flatiron Institute, 162 5th Ave, New York, NY 10010, USA}

\author{Francisco Villaescusa-Navarro}\affiliation{Department of Astrophysical Sciences, Princeton University, Peyton Hall, Princeton, NJ, 08544, USA}
\affiliation{Center for Computational Astrophysics, Flatiron Institute, 162 5th Ave, New York, NY 10010, USA}

\author{David N. Spergel}\affiliation{Center for Computational Astrophysics, Flatiron Institute, 162 5th Ave, New York, NY 10010, USA}\affiliation{Department of Astrophysical Sciences, Princeton University, Peyton Hall, Princeton, NJ, 08544, USA}
\begin{abstract}
Traditional large-scale models of reionization usually employ simple deterministic relations between halo mass and luminosity to predict how reionization proceeds. We here examine the impact on modelling reionization of using more detailed models for the ionizing sources as identified within the $100~{\rm Mpc/h}$ cosmological hydrodynamic simulation {\sc Simba}, coupled with post-processed radiative transfer.
Comparing with simple (one-to-one) models, the main difference of using {\sc Simba} sources is the scatter in the relation between dark matter halos and star formation, and hence ionizing emissivity. We find that, at the power spectrum level, the ionization morphology remains mostly unchanged, regardless of the variability in the number of sources or escape fraction.   In particular, the power spectrum shape remains unaffected and its amplitude changes slightly by less than 5-10\%, throughout reionization, depending on the scale and neutral fraction.
Our results show that simplified models of ionizing sources remain viable to efficiently model the structure of reionization on cosmological scales, although the precise progress of reionization requires accounting for the scatter induced by astrophysical effects.

\end{abstract}


\section{Introduction} \label{sec:intro}
Many upcoming surveys, such as the Square Kilometer Array ~\citep[SKA,][]{2013ExA....36..235M}, the Hydrogen Epoch of Reionization Array~\citep[HERA,][]{2017PASP..129d5001D}, the Low Frequency Array~\citep[LOFAR,][]{2013A&A...556A...2V}, are devoted to detecting reionization in the near future via its impact on HI 21cm emission topology. On the other hand, the James Webb
Space Telescope~\citep[JWST,][]{2006SSRv..123..485G} the Vera C. Rubin Observatory Legacy Survey of Space and Time~\citep[LSST,][]{2019ApJ...873..111I}, and Nancy Grace Roman Space Telescope~\citep[Roman,][]{2015arXiv150303757S}, are devoted to detecting the primary sources driving reionization at high-redshift (galaxies and quasars). With these growing observational efforts, we are moving towards an era of detailed observations of high-redshift galaxy formation. 
It is therefore important to test and quantify the uncertainties in our current models of reionization, by exploring the interplay between galaxy formation and cosmology, in order to prepare for extracting the maximum amount of information from the future surveys.

The main sources of uncertainties in reionization models arise from the ionizing source population \citep{2016ARA&A..54..761S}, the clumpiness of the ionized gas (which sets the number of recombinations; \citealt{2020ApJ...898..149D}), and approximations in performing the radiation transport (RT) \citep{2021JCAP...02..042W}. The impact of RT has been investigated in detail in several studies~\citep[e.g.][]{2014MNRAS.443.2843M,2019MNRAS.489.5594M,2011MNRAS.414..727Z,2011MNRAS.411..955M}, who found that, using the same ionizing sources, the Excursion-Set Formalism~\citep{1974ApJ...187..425P,1991ApJ...379..440B}, which is an algorithm to identify maximally ionized bubbles within spherical regions around sources, produces similar topology to the post-processing radiative transfer of cosmological hydrodynamic simulations. Hence, it appears justified to use excursion-set based models to study the reionization signal on large scales. Meanwhile, the impact of including inhomogeneous recombination and clumping effects --- the sink model --- has also been studied in several works~\citep[e.g.][]{2014MNRAS.440.1662S,Hassan:2016}, which concluded that inhomogeneous recombination reduces the ionized region size, delays reionization, and suppresses the large scale 21~cm power. In contrast, the {\it large scale} impact of including more physically motivated recipes to model the ionizing source population has not been studied.  This is the goal of this work.

In ~\citet{Hassan:2016}, we took a first step to study the impact of modelling ionizing sources more accurately within semi-numerical simulations, and showed that including power-law mass-dependent ionizing emissivity motivated from hydrodynamical simulations as opposed to the usual linear dependence on mass (through an efficiency parameter $\zeta\equiv  R_{\rm ion}/M_{\rm h}$ of ionizing photon emissivity \rion vs. halo mass $M_{\rm h}$) increases the duration of reionization and boosts the small scale 21cm power spectrum. This indicates that a modest change in the ionizing emissivity--halo mass (\rion-$M_{\rm h}$) relation has a major impact on reionization globally. This impact owes to the more massive sources in the super-linear scaling being more clustered and hence producing larger HII regions \citep{2006MNRAS.365..115F, 2007MNRAS.377.1043M}

Currently, the most accurate way to model ionizing sources in a cosmological volume is through state-of-the-art cosmological hydrodynamical galaxy formation simulations, in which star formation is modelled following, for instance,  the sub-grid multiphase model of ~\citet{2003MNRAS.339..289S} or the H$_{2}$-regulated model of~\citet{2009ApJ...693..216K}. Such simulations include {\sc Illustris}~\citep{2014MNRAS.445..175G}, {\sc IllustrisTNG }~\citep{2018MNRAS.473.4077P}, {\sc Eagle}~\citep{2015MNRAS.446..521S}, Horizon-AGN~\citep{2017MNRAS.467.4739K} \simba~\citep{2019MNRAS.486.2827D},  Cosmic Reionization on Computers ~\citep[CROC,][]{2014ApJ...793...29G,2014ApJ...793...30G}, Cosmic Dawn II~\citep[CoDA II,][]{2020MNRAS.496.4087O}, SPHINX~\citep[][]{2021MNRAS.507.1254K}, THESAN~\citep[][]{2022MNRAS.511.4005K} using the AREPO-RT~\citep[][]{2019MNRAS.485..117K}, to name a few. It has recently become possible to run galaxy formation simulations on cosmological scales ($\gtrsim 100\, {\rm Mpc}$) while resolving halos down to $M_{\rm h}<10^{10}$ M$_{\odot}$, thanks to advances in computing capabilities. Hence, it is now possible to ask the question: {\it Does detailed astrophysics matter for modeling cosmic reionization?}

The main difference in using sources identified within these galaxy formation simulations consists in including the scatter around the \rion--$M_{\rm h}$ relation. This scatter, which is due to stellar feedback, merger history  and/or photo escape fraction, might have a direct correlation with the density field and/or other properties such as metallicity, and hence its impact on large scale reionization remains non-trivial. As a simple example, by assuming a lognormal scatter around sources identified within  semi-numerical excursion-set based models,~\citet{2020MNRAS.495.3602W} showed that the scatter reduces the Ly$\alpha$ visibility for brightest galaxies. 

In this work, we use ionizing sources identified within the state-of-the-art \simba\ simulation, that has been shown to reproduce a wide range of observations, at both low-z~\citep[see][]{2019MNRAS.486.2827D} and during the reionization epoch~\citep{2020MNRAS.494.5636W,2020ApJ...905..102L}. We consider a broad range of source models that uses the SIMBA star formation rates as inputs, but considering different minimum halo masses for sources owing to resolution limits in the simulations and varied escape fraction pictures.  We then run large scale radiative transfer reionization simulations, with the radiative transfer done in post-processing using {\sc ATON}~\citep{Aubert:2008,Aubert:2010} via a moment-based method with an M1 closure relation.  We compare these results with a simple power-law model of the ionizing emissivity as a function of halo mass, as commonly used in large-scale semi-numerical models. Compared to previous works~\citep[e.g.][]{2020MNRAS.495.3602W}, our approach has several advantages and improvements, namely the scatter in ionization rate naturally emerges from the detailed sub-grid physics implemented within \simba, and the impact on large scale topology is determined through accurate modelling of radiative transfer with {\sc ATON}. 

This paper is organized as follows: we briefly describe the \simba\ simulation in \S~\ref{sec:sim}, introduce the source models in \S~\ref{sim:source}, explain how the reionization realization is generated in \S~\ref{sec:eor}, present our results in \S~\ref{sec:res}, and conclude in \S~\ref{sec:con}. Throughout this work, we adopt a $\Lambda$CDM cosmology with parameters $\Omega_{m}=0.3$, $\Omega_{\Lambda}=0.7$, $\Omega_{b}=0.048$, $h=0.68$, $\sigma_{8}=0.82$, and $n_{s}=0.97$, consistent with~\citet{2016A&A...594A..13P} constraints. All quantities are in comoving units unless otherwise noted.

\section{Methods}

\subsection{\simba\ Simulations}\label{sec:sim}

We here briefly describe {\sc Simba} and refer the reader to~\citet{2019MNRAS.486.2827D} for more detailed information about the physics implemented in the simulation.

The {\sc Simba} model was introduced in ~\citet{2019MNRAS.486.2827D}. {\sc Simba} is a follow-up to the {\sc Mufasa}~\citep{Dave:2016} cosmological galaxy formation simulation using {\sc Gizmo}'s meshless finite mass hydrodynamics solver~\citep{Hopkins:2015,Hopkins:2018}. Radiative cooling and photo-ionisation heating are implemented using the updated {\sc Grackle$-3.1$} library~\citep{Smith:2017}. A spatially-uniform ionizing background from~\citet{Haardt:2012} is assumed, in which self-shielding is accounted for following~\citet{Rahmati:2013}. The chemical enrichment model tracks nine elements (C,N,O,Ne,Mg,Si,S,Ca,Fe) arising from Type II supernovae (SNe), Type Ia SNe, and Asymptotic Giant Branch (AGB) stars. Type Ia SNe and AGB wind heating are also included. The star formation-driven galactic winds are kinetically launched and decoupled into hot and cold phase winds, and  the winds are metal-loaded owing to local enrichment from supernovae, with overall metal mass being conserved. The mass rate entering galactic outflows is modelled with a broken power law following~\citet{Angles:2017b}.  The quasi-linear scaling of wind velocity with escape velocity from~\citet{Muratov:2015} is adopted.  {\sc Simba} further implements black hole growth via a torque-limited accretion model~\citep{Angles:2017a}, and the
feedback from Active Galactic Nuclei (AGN) is implemented in both  radiative and jet modes. {\sc Simba} also accounts for the X-ray AGN feedback in the surrounding gas following~\citet{Choi:2012}. Dust production and destruction are modeled on-the-fly, leading to predictions of dust abundance that broadly agree with a variety of constraints over cosmic time~\citep{Li:2019}. 

{\sc Simba} employs a molecular gas-based prescription following~\citet[hereafter KMT]{Krumholz:2009} to form stars. KMT is a physically motivated recipe to model star formation as seen in local disk galaxies, where a strong correlation is seen between SFR and molecular gas content~\citep[e.g.][]{Leroy:2008}. Star formation is only allowed in the dense gas phase (ISM gas) above a hydrogen number density $n_{\rm H}$ $\geq$ 0.13 cm$^{-3}$, though in practice the H$_2$ fraction forces star formation to occur at much higher densities ($n_{\rm H}\gg 1$~cm$^{-3}$). {\sc Simba} assumes a ~\citet{Chabrier:2003} initial mass function (IMF) throughout.

{\sc Simba} is an attractive platform to use in this study due its ability to match a wide range of observations as show in ~\citet{2019MNRAS.486.2827D}. These observations include the stellar mass function evolution,  main sequence, evolution in gas fractions (both neutral and molecular), the gas–phase and stellar mass–metallicity relations, galaxy photometric projected sizes, hot halo gas fractions, black hole mass-stellar mass relation, and the dust mass function. On the other hand, {\sc Simba} fails to produce a sharp knee in the steller mass function at z=0, and produces larger low-mass quenched galaxy sizes, to name few examples. Despite these small disagreements, {\sc Simba} remains a state-of-the-art platform for studying the scatter impact on reionization morphology.
\subsection{Source Models}\label{sim:source}
The ionization rate ($R_{\rm ion}$) from each galaxy is computed using a stellar metallicity-dependent parameterization of the ionizing photon flux ($Q(Z)$), which reproduces the equilibrium values compiled by~\citet{2003A&A...397..527S}. This $Q(Z)$ parameterization is provided by \citet{2011ApJ...743..169F}:
\begin{equation}
        \log_{10} Q(Z) = 0.639(-\log_{10}Z)^{1/3} + 52.62 - 0.182\, ,
\end{equation}

where $Q$ is in units of s$^{-1} (M_{\odot} {\rm yr}^{-1})^{-1}$ and the last term converts to a~\citet{Chabrier:2003} IMF. It is straightforward to convert the $Q$ to $R_{\rm ion}$ using the SFR as follows:
\begin{equation}\label{eq:rion}
        R_{\rm ion} = Q \times {\rm SFR}.
\end{equation}

 We directly  compute the ionizing emissivity from the SIMBA halo catalogs. We use the stellar mass-weighted metallicity Z of all star particles and the total SFR of all gas particles within each halo to evaluate Equation~\ref{eq:rion} in order to provide the emissivity required for the radiative transfer calculations.
\begin{figure*}
    \centering
    \includegraphics[width=\textwidth]{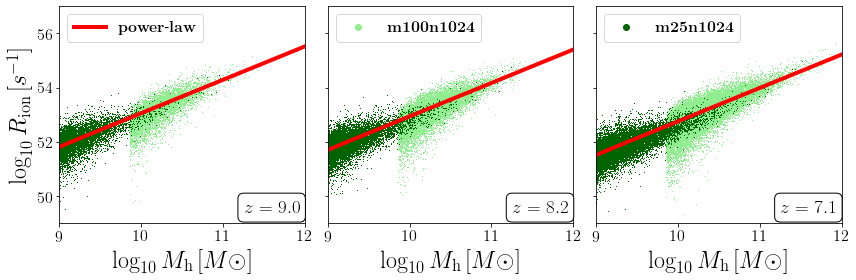}
    \caption{Ionization rate $R_{\rm ion}$ as a function of halo mass $M_{\rm h}$ from resolved halos at different redshifts in fiducial (m100n1024, light green) and high (m25n1024, dark green) resolution {\sc Simba} runs using Equation~\ref{eq:rion}. The red line represents the fitting functions (from Equation~\ref{eq:fitting}), which we will use later to study the scatter impact.}
   \label{fig:rion_fit}
\end{figure*}

\begin{figure}
    \centering
    \includegraphics[scale=0.45]{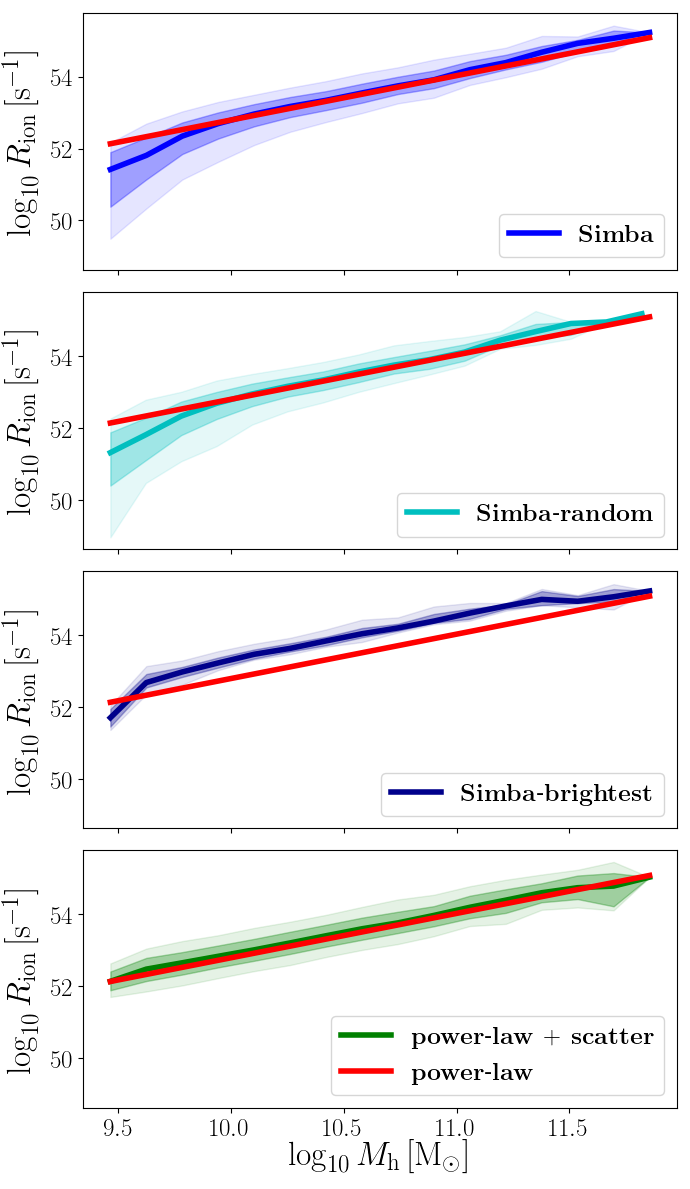}
    \caption{Shows the ionization rate $R_{\rm ion}$ as a function of halo mass at $z\sim$7 for the different source models considered in this study. The top three panels show different {\sc Simba} variants and the bottom panel shows the {\rm power-law} model with/without scatter.  In all panels, we show the {\bf power-law} in red. Solid lines represent the 50th percentile of the different $R_{\rm ion}$ distributions, and the dark and light shaded areas are the corresponding 1-2 $\sigma$ levels, respectively.}
   \label{fig:rion_all}
\end{figure}

 \begin{figure*}
    \centering
    \includegraphics[scale=0.8]{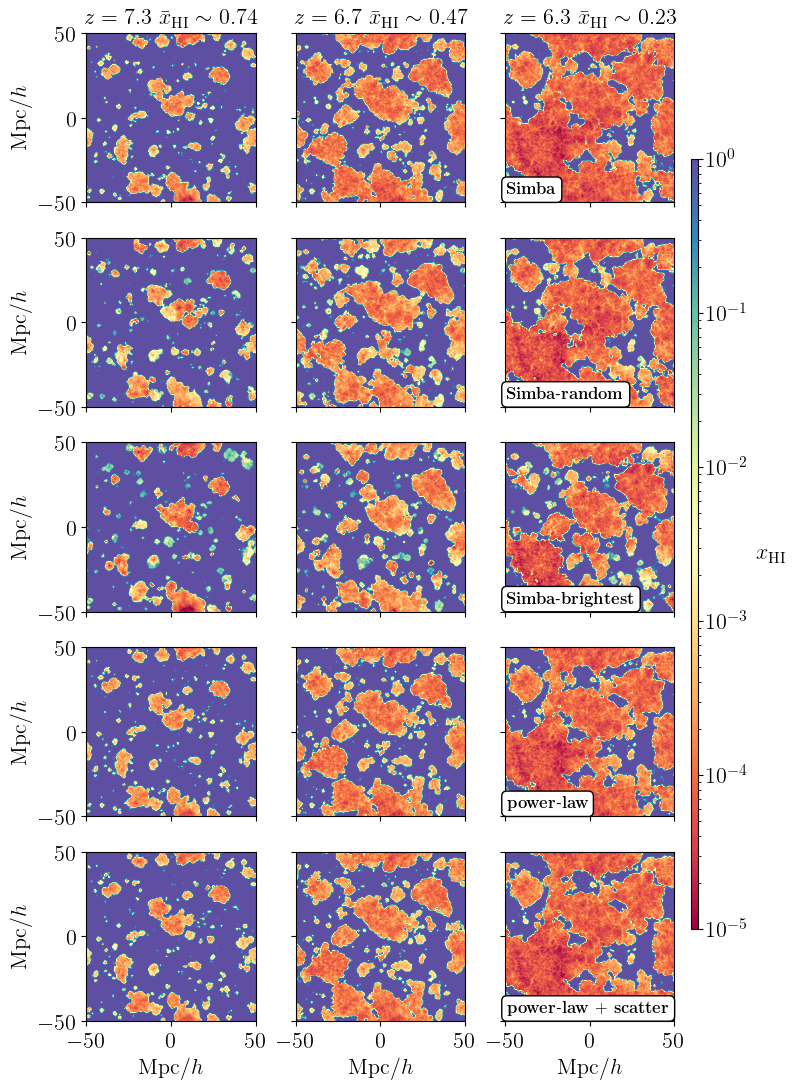}
    \caption{Comparison between ionization maps for all models, at different redshifts and stages during reionization. All models produce similar maps, indicating that the scatter, either in the ionizing emissivity ({\sc Simba} vs power-law models) or in the photon escape fractions (comparing all {\sc Simba} variants), has a minimal impact on the large scale morphology.   }
    \label{fig:maps}
\end{figure*}

As mentioned earlier, our aim is to use these realistic sources, as identified within {\sc Simba}, to study the impact on reionization on cosmological scales, as compared to simplified models. We use the largest {\sc Simba} run available which has a volume of 100$h^{-1}$ Mpc with  1024$^{3}$ dark matter and gas particles each. While this simulation has a sufficiently large volume suitable to study large scale reionization~\citep{2014MNRAS.439..725I}, the dark matter particle mass resolution is about 9.6$\times$10$^{7}$ $M_{\odot}$, and hence the minimum halo mass resolved would be of order 10$^{9} M_{\odot}$ assuming 16 particles as minimum to identify halos in the friends-of-friends finder.  However, if the resolution is high enough to resolve low mass halos down to 10$^{8} M_{\odot}$, the number of sources increases, and in this case the scatter impact would be even smaller since the ionizing emissivity average in pixels is now taken over a larger number of sources. Later we include models with different number of sources to address this point.  

To construct a simple model of the average  $R_{\rm ion}$ as a function of halo mass -- a comparison model that averages out the fluctuations of individual halos -- we follow our earlier approach presented in~\citet{Hassan:2016} where we obtained a parameterization for the ionizing emissivity $R_{\rm ion}$ as a function of halo mass $M_{\rm h}$ and redshift $z$ from hydrodynamic simulations. We here repeat the same exercise to obtain the best fitting parameters for the following simple model of $R_{\rm ion}$:

\begin{equation}\label{eq:fitting}
    R_{\rm ion} / M_{h} = A(1+z)^{D} \left( M_{h}/B \right)^{C} \exp\left( - (B/M_{h})^{3} \right), 
\end{equation}

where $A,B, C$ and $D$ are free parameters. This fitting function, which is similar to the Schechter function, is motivated by the fact the $R_{\rm ion}$ behaves as a power-law at high halo masses and exhibits a turnover at the low mass end. In~\citet{Hassan:2016}, we found the turnover occurs around 10$^{8} M_{\odot}$ (see Figure 1 therein). Since the largest \simba\ run used in this study resolves halos down to $\sim 10^{9-10} M_{\odot}$, we refer to models using this simple relation (Equation \ref{eq:fitting}) as {\it power-law}.    

In order to estimate the slope of $R_{\rm ion}$ more accurately from \simba, we include a higher resolution run which has a smaller volume of 25$h^{-1}$Mpc with the same number of particles ($2\times1024^{3}$). Hence, this run resolves halos down to 10$^{8.5} M_{\odot}$; halos below this mass are not expected to contribute significantly~\citep{Finlator:2011}. We combine these two simulations at different redshifts ($z=6-10$) to obtain the $R_{\rm ion}$ best fitting parameters as shown in Figure~\ref{fig:rion_fit}. Conservatively, one may obtain the best-fit parameters using catalogs from high (e.g. z=20) to low (e.g. z=6) redshifts. However, restricting the fitting to redshifts lower than z=10 would not impact the results since reionization, in models considered in this study, begins only afterward.  This figure shows $R_{\rm ion}$ as a function of $M_{\rm h}$ at different $z$ from \simba\ m100n1024 (light green) and m25n1024 (dark green) runs using Equation~\ref{eq:rion}. The fitting function, Equation~\ref{eq:fitting}, is shown in red, and the best fitting parameters are found using a non-linear least square fit as follows: $A=3.51\times 10^{39} \pm 5.6\times 10^{37} [\rm sec^{-1}]$,  $B = 3.3\times 10^{8} \pm 2.6\times 10^{6}  [M_{\odot}]$, $C=0.234 \pm 0.00078$ and $D=3.2 \pm 0.0082$. These parameters are similar to those found earlier by~\citet{Hassan:2016}, based on full radiative hydrodynamic simulations (with much smaller volume). It is evident that these parameters produce a good fit to the ionizing emissivity at different redshifts, and hence we will fit this as our power-law model of $R_{\rm ion}$ to compare with using a more realistic $R_{\rm ion}$ directly from \simba.

To perform a consistent comparison and to carefully investigate the scatter impact, we tune the photon escape fraction so that all models would produce roughly the same number of ionizing photons. In addition, we ensure that all  models do produce the expected ionizing emissivity evolution~\citep[e.g.][]{Becker:2013b} as well as the neutral fraction measurements by the end of reionization~\citep[e.g.][]{2006AJ....132..117F,2015ApJ...811..140B}. It is worth noting that, for the purpose of this study, it is not important to match or reproduce any measurements. The most important part of this analysis is to ensure that the comparison is performed at the same redshift and neutral fraction in order to exclude any other cosmological and astrophysical effects.

Using the same density field and recombination rate, we consider five different source models, that are tuned to emit similar amount of ionizing photons and obtain same reionization history as follows:

\begin{itemize}
    \item {\bf Simba}: uses all sources with emissivities computed using Equation~\ref{eq:rion}, and $f_{\rm esc}=0.25$ (blue in Figure~\ref{fig:rion_all}). 
    \item {\bf Simba-random}: uses only 25\% of all sources, randomly selected within different halo mass bin of 0.25 dex. Similar to {\bf Simba}, the emissivity is computed using Equation~\ref{eq:rion}, but with $f_{\rm esc}=1.0$ (cyan in Figure~\ref{fig:rion_all}).
    \item {\bf Simba-brightest}: uses only the brightest halos  (the most luminous halos) in each mass bin of 0.25 dex (halos with the highest $R_{\rm ion}$). These halos are about 7\% out of all halos. Similar to {\bf Simba-random} in terms of emissivity and unity $f_{\rm esc}$ (dark blue in Figure~\ref{fig:rion_all}). 
    \item {\bf power-law}: uses all sources with emissivities computed using Equation~\ref{eq:fitting}, and $f_{\rm esc}=0.25$ (red line in Figure~\ref{fig:rion_all}). 
    \item {\bf power-law with scatter}: Same as {\bf power-law}, except with the addition of scatter assuming that \rion may be represented by a lognormal distribution with a mean given by Equation~\ref{eq:fitting} and standard deviation set to 0.3 dex~\footnote{This amount of scatter (0.3 dex) is the average scatter over the full halo mass range as measured in \simba.}. This model uses $f_{\rm esc}=0.2$ (green in Figure~\ref{fig:rion_all}).
\end{itemize}

A visual summary of these different models is provided in Figure~\ref{fig:rion_all}, where solid lines represent the 50th percentile of the different $R_{\rm ion}$ distributions, and the dark and light shaded areas are the corresponding 1-2 $\sigma$ levels, respectively. As a reference, we show in red the {\bf power-law} model in all panels.  All these models include different sources of scatter.  For instance, the scatter in {\bf Simba-random} and {\bf Simba-brightest} models is based on the  variability of the photon escape fraction, and consider an extreme case where some halos (whether randomly or the brightest) have $f_{\rm esc}=1$ and the rest are completely obscured ($f_{\rm esc}=0$). The {\bf Simba-brightest} is motivated by the fact that galaxies with high SFR would have stronger feedback (more gas ejection) and hence higher $f_{\rm esc}$. The other models, {\bf Simba} or {\bf power-law with scatter}, include different forms of scatter either from {\sc Simba}'s detailed sub-grid physics or by assuming a simple lognormal distribution, respectively, while assuming the same $f_{\rm esc}$. It is worth noting that the {\bf Simba} and {\bf power-law} models are found to obtain same reionization history using the same photon escape fraction of 25\%. This shows that the {\bf power-law} model is a good representative of the average $R_{\rm ion}$ in {\sc Simba}.  In fact, the {\bf power-law} approximately represents the average of all considered models, expect the {\bf Simba-brightest} where the models differ by a factor of 2-3. This shows that comparisons with these different models would reveal the impact of variability within either the number of sources or $f_{\rm esc}$. While the scatter in {\sc Simba} increases the total $R_{\rm ion}$ by $\sim 20 \%$  at the massive end, this increase has a negligible effect on the reionization history since the number density of massive sources ($> 10^{11} M_{\odot}$) is small as compared to the faint ones. In addition, it is also seen in the top panel of Figure~\ref{fig:rion_all} that the {\bf power-law} has a higher $R_{\rm ion}$ slope at low mass end than the average in {\sc Simba} (blue solid line), which boosts the total $R_{\rm ion}$ in the {\bf power-law} model to a similar level as compared to {\sc Simba}. By contrast, the {\bf power-law  with scatter} requires a 20 \% lower $f_{\rm esc}$ as compared to the {\bf power-law}. This is a consequence of assuming that the $R_{\rm ion}$ can be modelled using a log-normal distribution, where the median is somewhat less than  the mean. This shows that $f_{\rm esc}$ is highly sensitive to the assumptions used to build the source model. We compare all these models in terms of their large scale ionization  morphology in \S\ref{sec:res}.

\begin{figure*}[htp!]
    \centering
    \includegraphics[scale=0.46]{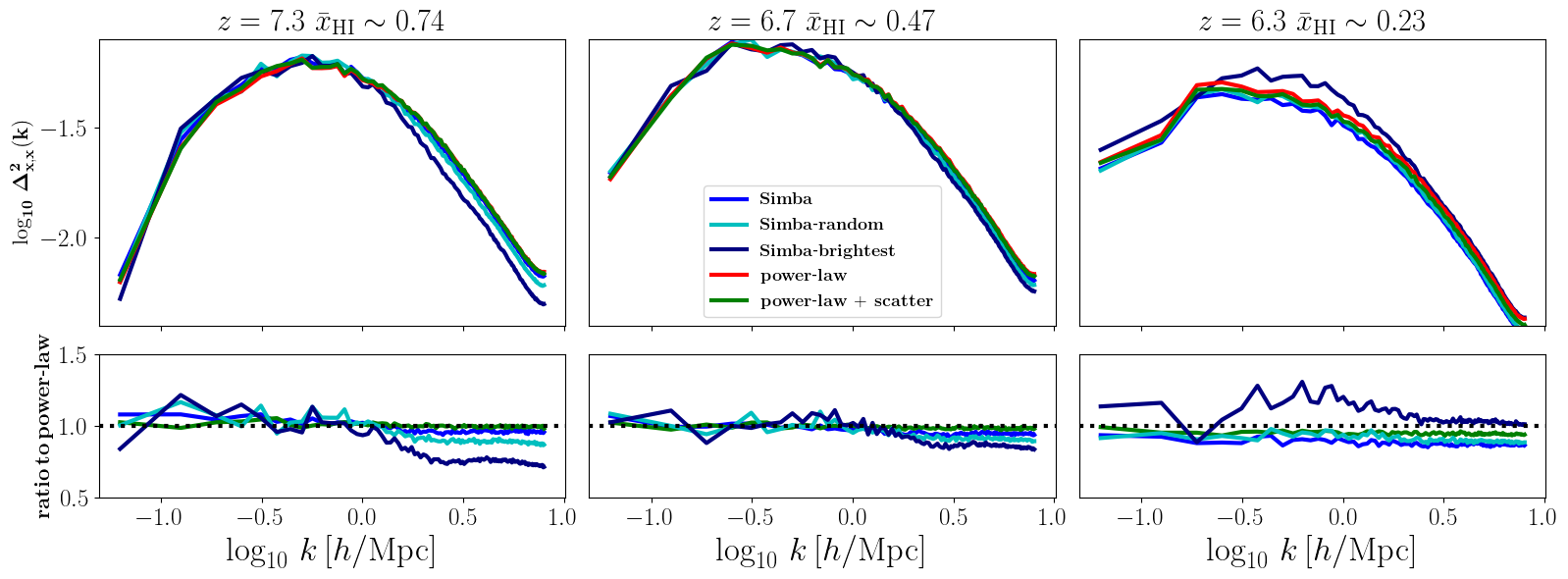}
    \caption{Comparison between ionization power spectra for all models, at different redshifts and stages during reionization (top panels). The bottom panels show the power ratio of all models to the power-law model. As seen in Figure~\ref{fig:maps}, all models produce similar power spectra on small and large scales, irrespective of the variability in the emissivity or in the photon escape fraction. }
    \label{fig:power}
\end{figure*}

\subsection{Radiative Transfer}\label{sec:eor}
All reionization realizations from different source models are obtained in post-processing using the radiative transfer module {\sc ATON}~\citep{Aubert:2008,Aubert:2010}, which is a moment based method with M1 closure relation. We run reionization from $z\sim 20$ down to $z\sim5$, using all snapshots available in this redshift range, which are about 35 snapshots in the time interval of about 20 Myr. Single frequency has been used, and temperature has been evolved. We assume a black body spectrum with temperature $T=5\times 10^{4} K$ (E=20.27 eV), and full speed of light (c=1). A similar version of {\sc ATON} has been used in~\citet[e.g.][]{2020MNRAS.491.1736K} and \citet{2019MNRAS.485L..24K}, for instance, to study the large scale opacity fluctuations and Lyman-$\alpha$ forest by the end of reionization. The main difference in our analysis lies in considering different source models as explained in the previous section.  We first smooth densities of gas particles on a grid with number of cells N=256$^{3}$, resulting in a resolution of $390.625 \, {\rm ckpc}/h$.  While this is somewhat coarse density grid, in~\citet{Hassan:2016}, we have performed a convergence test using the same source model \rion for different resolutions and configurations, and found that the change in the power spectrum remains minimal. Hence, we do not expect the number of cells would alter the conclusion. Using the gas density grids with the ionizing emissivities as inputs, we use {\sc ATON} to self-consistently compute the ionization field, gas temperature and the photoionization rate. We run {\rm ATON} with all source models, and compare the results in the following section. 

\section{results: Reionization Morphology}\label{sec:res}
We here compare the effect of using the different source models described in the previous sections on reionization morphology. The main difference between these models lies in the scatter in the \rion-$M_{\rm h}$ relation, and hence this study shows the impact of including different forms of scatter on reionization. 

In Figure~\ref{fig:maps} and~\ref{fig:power}, we show a comparison between all models in terms of maps of the ionization field and their corresponding power spectra at different redshifts, respectively. We can see that the morphology remains mostly unaffected as seen in the maps as well as the power spectra. This is also seen in the ratio between all models to the power-law model (bottom panels in Figure~\ref{fig:power}), which is close to unity in all cases. However, it is expected for the {\bf Simba-random} and {\bf Simba-brightest} models to have a lower small scale power, due to the smaller number of sources as compared to other models. This effect is clearly seen at the early stages of reionization where bubbles are small. As reionization proceeds and bubbles grow, all models seem to have the same power ratio, except for the {\bf Simba-brightest} model, due to the difference in the global neutral fraction. In addition, the source clustering and bias only change in the {\bf Simba-brightest} model, which is the reason for its higher large scale power as compared to other models, particularly at the late stages of reionization. All these effects are too small to cause noticeable differences in the maps. It is worth mentioning that we have also checked the opposite scenario to the {\bf Simba-brightest} model, where only the faintest halos are used with unity $f_{\rm esc}$. In that case (i.e. {\bf Simba-faintest}), nearly 50\% of halos are required to yield similar emissivity and reionization history to other models. All results remain unchanged whether random, brightest or faintest halos are used. This minimal impact might be due to the fact that hundreds if not thousands of galaxies exist within large ionized bubbles, and hence the scatter would average out. In this case, the scatter impact would be reduced by a factor of  $\sim$ $1/\sqrt{N}$, where $N$ is the number of sources. Since these comparisons are performed at the same redshift and the 21cm brightness temperature field traces the ionization topology, the same conclusion is valid for the 21cm fluctuations on large scales. This shows that including different forms of scatter in the \rion-$M_{\rm h}$ relation has a minimal impact on reionization morphology. However, we have checked the bi-spectra between these models and found bigger differences (scatter induces a more negative bi-spectrum, hence larger skewness). We leave investigating the impact of scatter on higher order summary statistics to future works.

\section{conclusions}\label{sec:con}
We have studied the impact of scatter on reionization morphology by considering a broad range of models with different variations in the photon escape fractions and emissivities, using sources identified within the state-of-the-art galaxy formation simulation \simba\,. The main improvement over previous works is the inclusion of a realistic scatter in the \rion-$M_{\rm h}$ relation.  The radiative transfer is performed using the {\rm ATON} routine, which implements a moment based method with M1 closure relation. We post-process the radiative transfer on \simba's largest available simulation run, which has a comoving volume of $100\, {\rm Mpc}/h$, sufficient to yield convergent reionization histories and capture the bubble overlap epoch by the end of reionization~\citep{2014MNRAS.439..725I}. 

Besides models with scatter, we constructed a simple source model ({\bf power-law}) using a fitting function (Equation~\ref{eq:fitting}) that behaves as a power-law for halos larger than \simba's resolution limit ($ > 10^{9} M_{\odot}$). This model adopts a one-to-one \rion-$M_{\rm h}$ relation. We show in Appendix~\ref{app} that those missing halos below $10^{9} M_{\odot}$ 
do not change the reionization morphology but are important to accurately estimate the ionization history and astrophysical parameters such as the photon escape fraction. 

To investigate the impact of the scatter, we tune the number of sources and photon escape fractions in all models to produce similar evolution in the comoving ionizing emissivity density and reionization history. Our main key finding can be summarized as follows: {\it The scatter in the \rion-$M_{\rm h}$ relation does not significantly alter the reionization morphology as seen in the ionization maps (see Figure~\ref{fig:maps}) or the power spectra (see Figure~\ref{fig:power}) if computed at the same global ionization fraction.}

It is worthwhile to mention several limitations associated with this work. The radiative transfer is modelled using a moment based method with M1 closure that is known to over-ionize Lyman Limit Systems in post-reionization  and  does  not  accurately  capture  shadowing~\citep[e.g.][]{2021JCAP...02..042W}. The radiative transfer also has been run in post-processing and hence these results do not include the coupling between the hydrodynamics and radiative transfer. We also do not account for contribution from binary stars that would induce earlier reionization~\citep[e.g.][]{2018MNRAS.479..994R,2021MNRAS.505.2207D}.      Additionally, due to the low resolution in the current \simba\ run, we have relaxed the requirement to identify dark matter halos, assuming only 16 particles as a minimum to capture smaller halos. While we have checked explicitly that the results remain the same when we require a larger number of dark matter particles (e.g. 64) to identify halos, the comparison between {\bf Simba-random} and {\bf Simba-brightest} with {\bf Simba}, where a smaller number of sources were used (e.g. 25\% and 7\%), provides an evidence that, irrespective of the number of sources, models (with/without scatter) at approximately the same neutral fraction, would produce similar reionization morphology on cosmological scales. Recent works~\citep[e.g.][]{2020MNRAS.498.2001M,2021arXiv210501663O} have shown that the photon escape fraction depends on halo mass, and varies by $\sim$ 2 orders of magnitude between low and high mass halos. Although most of our models adopt a constant $f_{\rm esc}$, {\bf Simba-brightest} adopts an extreme case in $f_{\rm esc}$ variability, where 7\% of all halos (the brightest) are given unity $f_{\rm esc}$, and the rest (93\%) have $f_{\rm esc}=0$. This implies that, regardless of $f_{\rm esc}$ variability with halo mass, the reionization morphology remains relatively similar (comparing {\bf Simba-brightest} to other models). In Appendix~\ref{app2}, we attempt to gain deeper insights on the minimal impact of scatter on the large scale morphology by comparing the ionizing emissivity history for randomly selected halos in terms of the dark matter halo evolution versus the star formation history. 

 ~\citet[][]{2020MNRAS.498..430I} have performed a detailed comparison between different galaxy formation models (Illustris, IllustrisTNG, Mufasa, Simba, EAGLE), and found that the variability of star formation histories within these simulations broadly show a similar power spectral density that is well described by a broken power-law. This indicates that similar results would be obtained using any of these galaxy formation models, since the \rion history follows closely the SF history. In addition, most of radiative transfer hydrodynamic simulations use either the OTVET or M1 approximation to solve the moment-based radiative transfrer eqaution, which have been shown to perform similar during the early and intermediate stages of reionization, but  over-ionize self-shielding absorbers by the same amount by end of reionization~\citep[see e.g.][]{2021JCAP...02..042W}. This suggests that regardless of the radiative transfer method, similar conclusion might be drawn. As seen in Figure~\ref{fig:rion_fit}, the scatter decreases towards higher masses, which can be resolved within larger box sizes. Since the massive halos are very rare, we do not expect their presence to have a significant impact on the results but a slight modification to $f_{\rm esc}$ might be required to obtain the same reionization history.   It is worth mentioning that the scatter would increase the number of sources above a specific luminosity threshold, and hence future surveys with JWST surveys will be able to place tighter constrains on the scatter. We leave performing detailed JWST forecasting using {\sc Simba} to future works. 

Our results suggest that simplified models of ionizing sources, with some calibration of the parameters such as the photon escape fraction, can be used as a viable computationally efficient tool to model reionization on cosmological scales.

\section*{Acknowledgements} 
The authors acknowledge helpful discussions with Adam Lidz and Enrico Garaldi. We thank Dominique Aubert for  providing ATON, and Jonathan Chardin for guidance on  how to use it.  We thank the anonymous referee for their constructive comments which have improved the manuscript significantly. SH and RSS acknowledge support from the Simons Foundation.  RD acknowledges support from the Wolfson Research Merit Award program of the U.K. Royal Society. DAA acknowledges support by NSF grant AST-2009687. FVN acknowledges funding from the WFIRST program through NNG26PJ30C and NNN12AA01C.  LCK was supported by the European Union’s Horizon 2020 research and innovation programme under the Marie Skłodowska-Curie grant agreement No. 885990. \simba\ was run using the DiRAC@Durham facility managed by the Institute for Computational Cosmology on behalf of the U.K. STFC DiRAC HPC Facility. The equipment was funded by BEIS capital funding via STFC capital grants ST/P002293/1, ST/R002371/1 and ST/S002502/1, Durham University and STFC operations grant ST/R000832/1.

\appendix
\section{Contribution of halos $< 10^{9} M_{\odot}$}\label{app}
\begin{figure*}
    \centering
    \subfloat[Ionizing emissivity density, $\dot{N}_{\rm ion}$, evolution as a function of redshift.]{
    \includegraphics[width=0.4\textwidth]{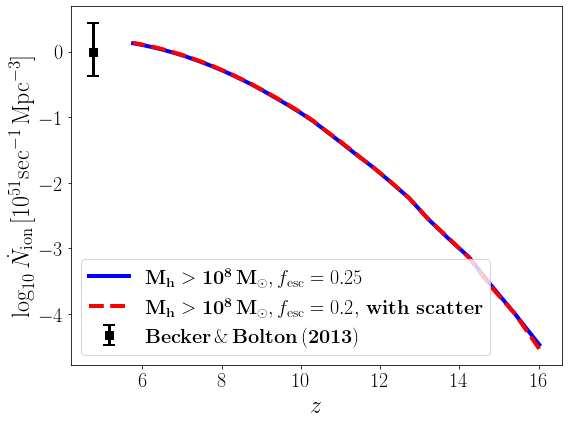}}
    \hspace*{\fill}
    \subfloat[Reionization history as compared with various measurements.]{
    \includegraphics[width=0.4\textwidth]{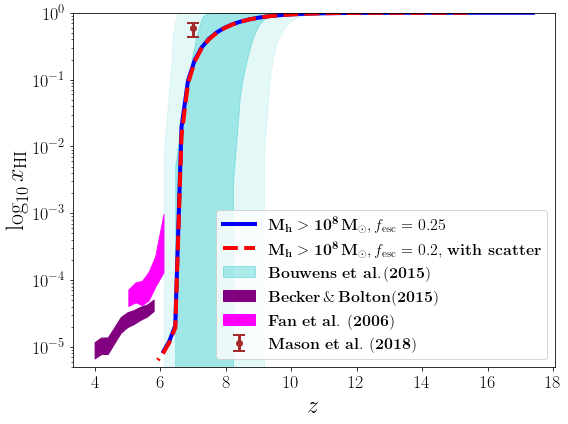}}
    \par
    \subfloat[Ionization maps and power spectra at different redshifts.]{
    \includegraphics[scale=0.4]{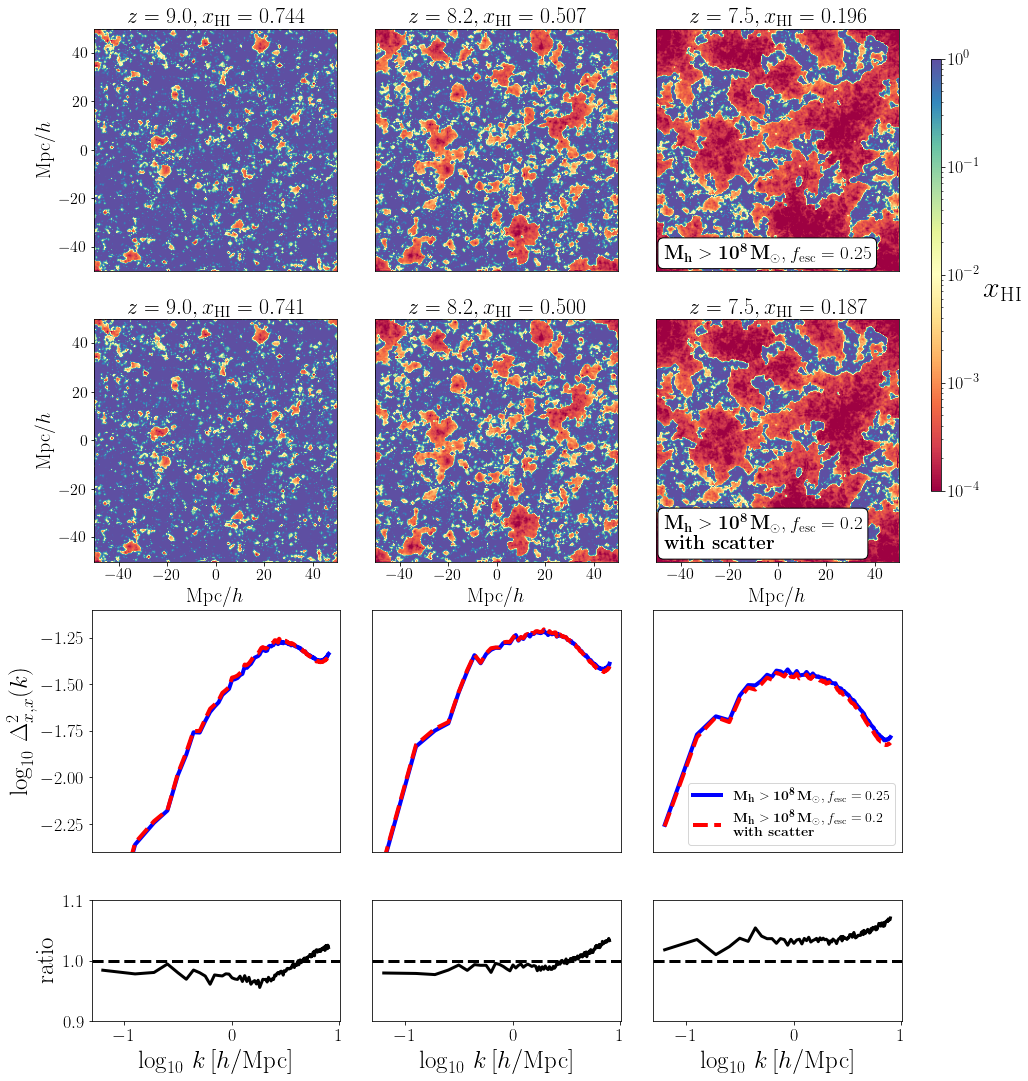}}
    \caption{Test with {\sc SimFast21} to examine the contribution of halos with $ M_{h} < 10^{9} M_{\odot}$ that are missing in {\sc Simba} due to resolution. This figure indicates that those halos have no impact on the large scale morphology.}
   
    \label{fig:globfig}
\end{figure*}

Since our results are based on the \simba\ simulation, which does not resolve halos below $< 10^{9} M_{\odot}$, we here aim to quantify whether these halos are expected to change the reionization signal on cosmological scales. For this test, we make use of a semi-numerical approach, {\sc SimFast21}~\citep{2010MNRAS.406.2421S}. We use the same cosmology, volume and number of cells, to generate the density field grids and halos using {\sc SimFast21}. We set the minimum halo mass in the excursion set formalism to $M_{h}=10^{8} M_{\odot}$. We then consider the {\bf power-law} model with/without scatter to run {\rm ATON} with the whole population of halos above $> 10^{8} M_{\odot}$, and compare results in Figure~\ref{fig:globfig}. We compare these two runs in terms of the reionization topology at relatively the same ionized fraction and redshift. We find that the topology remains unaffected as seen in the ionization maps and in the corresponding ionization power.  This indicates that the results presented in the main text remain valid, irrespective of these missing halos ($< 10^{9} M_{\odot}$).

\section{Variability in \rion history}\label{app2}
\begin{figure*}
    \centering
    \includegraphics[height=0.18\textheight]{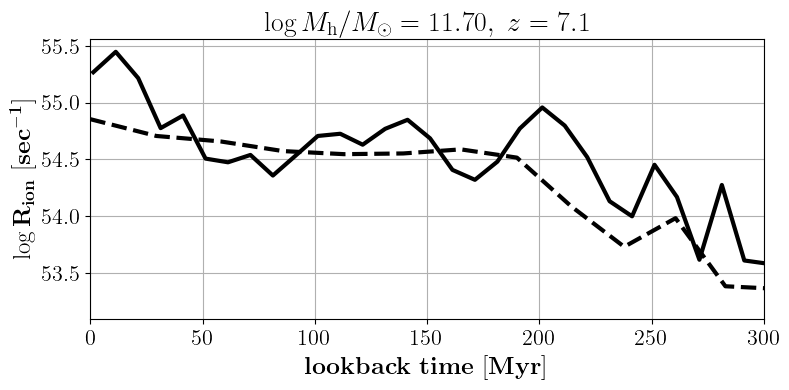}
    \includegraphics[height=0.18\textheight]{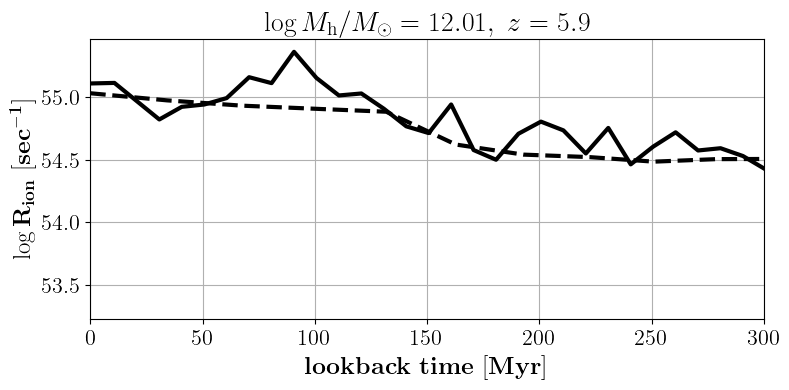}
    \includegraphics[height=0.18\textheight]{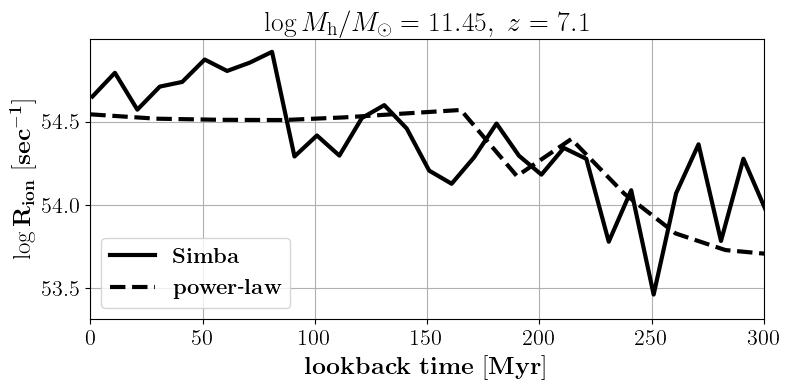}
    \includegraphics[height=0.18\textheight]{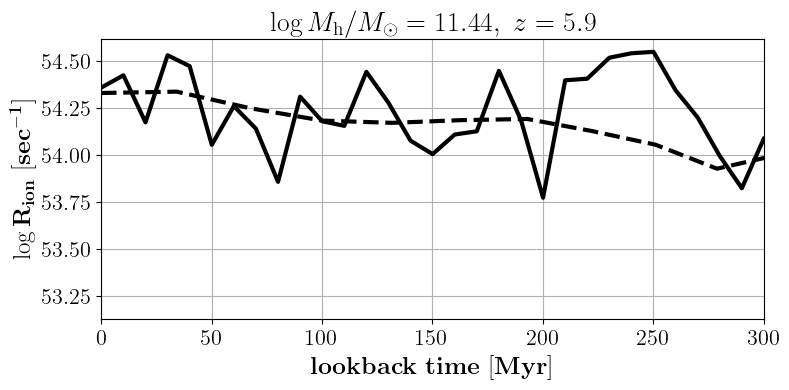}

    \caption{Examples of the $ \log_{10} R_{\rm ion}$ history for randomly selected halos as a function of lookback time using the star formation history ({\sc Simba}, solid) and the $R_{\rm ion}$ fitting formula (power-law, dashed) in 10 Myr intervals. Halo masses and redshifts are quoted in the figure legends above the individual panels. The redshift quoted above represents the zero point in the x-axis. The value of $R_{\rm ion}$ in {\sc Simba} varies by $\lesssim 1$ order of magnitude over time. The smooth evolution in the power-law model, which traces the evolution in dark matter halo mass, is a good approximation to the average $R_{\rm ion}$ evolution in {\sc Simba}. In all cases, the difference in instantaneous $R_{\rm ion}$  between these two models is minimal, and hence similar amounts of photons are emitted. This figure provides insights on why the impact of star formation variability on the global reionization signal is found to be minimal.   }
    \label{fig:rion_hist}
\end{figure*}

 To gain deeper insights on the scatter impact due to the evolution in dark matter or SF, we here focus on the variability in $R_{\rm ion}$ history for individual halos as shown in Figure~\ref{fig:rion_hist}. We first compute the $R_{\rm ion}$ history (solid lines) from {\sc Simba} using the star formation history, which is a measure of the total star formation occurring over a time interval ($\Delta t$), and the metallicity history, which quantifies the mass-weighted metallicity over $\Delta t$. We use the ages, masses and metallicities of all stellar particles associated with halos to compute these histories with $\Delta t=10 \,{\rm Myr}$, and the $R_{\rm ion}$ history is then computed using Equation~\ref{eq:rion}. We next compute the $R_{\rm ion}$ history differently in terms of the dark matter evolution over time  (dashed lines), in which we trace halos back in time to identify their progenitors and use their masses and redshifts in the power-law formula (Equation~\ref{eq:fitting}) to compute the $R_{\rm ion}$. We compare these different forms of $R_{\rm ion}$ histories in Figure~\ref{fig:rion_hist} for randomly selected halos, and quote their masses and redshift in the figure legend. The $R_{\rm ion}$ from {\sc Simba} (solid), which traces the star formation history, fluctuates by  $\lesssim 1$ order of magnitude over intervals of 10 Myr. On the other hand, the smooth evolution in the power-law model, which traces the evolution in dark matter (dashed lines), is a good approximation of the average $R_{\rm ion}$ evolution in {\sc Simba}.  In all cases, the difference in the instantaneous $R_{\rm ion}$ between these different $R_{\rm ion}$ histories is small, and hence similar amount of photons are emitted in each $R_{\rm ion}$ model. If we instead use a larger bin size in time (e.g. $\Delta t=30\, {\rm or}\, 40$ Myr), then the star formation history becomes more smooth and similar to the dark matter evolution as seen in the power-law model. This indicates that the scatter might have a minimal impact on the large scale reionization signal, since the average of $R_{\rm ion}$ over time is approximately similar to the smooth evolution of $R_{\rm ion}$ as computed from the evolution in dark matter. This exercise provides insights on why the impact of star formation variability on the global reionization signal is found to be minimal.  

\bibliography{sample63}{}

\begin{thebibliography}{}
\expandafter\ifx\csname natexlab\endcsname\relax\def\natexlab#1{#1}\fi
\providecommand{\url}[1]{\href{#1}{#1}}
\providecommand{\dodoi}[1]{doi:~\href{http://doi.org/#1}{\nolinkurl{#1}}}
\providecommand{\doeprint}[1]{\href{http://ascl.net/#1}{\nolinkurl{http://ascl.net/#1}}}
\providecommand{\doarXiv}[1]{\href{https://arxiv.org/abs/#1}{\nolinkurl{https://arxiv.org/abs/#1}}}

\bibitem[{{Angl{\'e}s-Alc{\'a}zar}
  {et~al.}(2017{\natexlab{a}}){Angl{\'e}s-Alc{\'a}zar}, {Dav{\'e}},
  {Faucher-Gigu{\`e}re}, {{\"O}zel}, \& {Hopkins}}]{Angles:2017a}
{Angl{\'e}s-Alc{\'a}zar}, D., {Dav{\'e}}, R., {Faucher-Gigu{\`e}re}, C.-A.,
  {{\"O}zel}, F., \& {Hopkins}, P.~F. 2017{\natexlab{a}}, \mnras, 464, 2840,
  \dodoi{10.1093/mnras/stw2565}

\bibitem[{{Angl{\'e}s-Alc{\'a}zar}
  {et~al.}(2017{\natexlab{b}}){Angl{\'e}s-Alc{\'a}zar}, {Faucher-Gigu{\`e}re},
  {Kere{\v s}}, {Hopkins}, {Quataert}, \& {Murray}}]{Angles:2017b}
{Angl{\'e}s-Alc{\'a}zar}, D., {Faucher-Gigu{\`e}re}, C.-A., {Kere{\v s}}, D.,
  {et~al.} 2017{\natexlab{b}}, \mnras, 470, 4698, \dodoi{10.1093/mnras/stx1517}

\bibitem[{{Aubert} \& {Teyssier}(2008)}]{Aubert:2008}
{Aubert}, D., \& {Teyssier}, R. 2008, \mnras, 387, 295,
  \dodoi{10.1111/j.1365-2966.2008.13223.x}

\bibitem[{{Aubert} \& {Teyssier}(2010)}]{Aubert:2010}
---. 2010, \apj, 724, 244, \dodoi{10.1088/0004-637X/724/1/244}

\bibitem[{{Becker} \& {Bolton}(2013)}]{Becker:2013b}
{Becker}, G.~D., \& {Bolton}, J.~S. 2013, \mnras, 436, 1023,
  \dodoi{10.1093/mnras/stt1610}

\bibitem[{{Bond} {et~al.}(1991){Bond}, {Cole}, {Efstathiou}, \&
  {Kaiser}}]{1991ApJ...379..440B}
{Bond}, J.~R., {Cole}, S., {Efstathiou}, G., \& {Kaiser}, N. 1991, \apj, 379,
  440, \dodoi{10.1086/170520}

\bibitem[{{Bouwens} {et~al.}(2015){Bouwens}, {Illingworth}, {Oesch}, {Caruana},
  {Holwerda}, {Smit}, \& {Wilkins}}]{2015ApJ...811..140B}
{Bouwens}, R.~J., {Illingworth}, G.~D., {Oesch}, P.~A., {et~al.} 2015, \apj,
  811, 140, \dodoi{10.1088/0004-637X/811/2/140}

\bibitem[{{Chabrier}(2003)}]{Chabrier:2003}
{Chabrier}, G. 2003, \pasp, 115, 763, \dodoi{10.1086/376392}

\bibitem[{{Choi} {et~al.}(2012){Choi}, {Ostriker}, {Naab}, \&
  {Johansson}}]{Choi:2012}
{Choi}, E., {Ostriker}, J.~P., {Naab}, T., \& {Johansson}, P.~H. 2012, \apj,
  754, 125, \dodoi{10.1088/0004-637X/754/2/125}

\bibitem[{{D'Aloisio} {et~al.}(2020){D'Aloisio}, {McQuinn}, {Trac}, {Cain}, \&
  {Mesinger}}]{2020ApJ...898..149D}
{D'Aloisio}, A., {McQuinn}, M., {Trac}, H., {Cain}, C., \& {Mesinger}, A. 2020,
  \apj, 898, 149, \dodoi{10.3847/1538-4357/ab9f2f}

\bibitem[{{Dav{\'e}} {et~al.}(2019){Dav{\'e}}, {Angl{\'e}s-Alc{\'a}zar},
  {Narayanan}, {Li}, {Rafieferantsoa}, \& {Appleby}}]{2019MNRAS.486.2827D}
{Dav{\'e}}, R., {Angl{\'e}s-Alc{\'a}zar}, D., {Narayanan}, D., {et~al.} 2019,
  \mnras, 486, 2827, \dodoi{10.1093/mnras/stz937}

\bibitem[{{Dav{\'e}} {et~al.}(2016){Dav{\'e}}, {Thompson}, \&
  {Hopkins}}]{Dave:2016}
{Dav{\'e}}, R., {Thompson}, R., \& {Hopkins}, P.~F. 2016, \mnras, 462, 3265,
  \dodoi{10.1093/mnras/stw1862}

\bibitem[{{DeBoer} {et~al.}(2017){DeBoer}, {Parsons}, {Aguirre}, {Alexander},
  {Ali}, {Beardsley}, {Bernardi}, {Bowman}, {Bradley}, {Carilli}, {Cheng}, {de
  Lera Acedo}, {Dillon}, {Ewall-Wice}, {Fadana}, {Fagnoni}, {Fritz},
  {Furlanetto}, {Glendenning}, {Greig}, {Grobbelaar}, {Hazelton}, {Hewitt},
  {Hickish}, {Jacobs}, {Julius}, {Kariseb}, {Kohn}, {Lekalake}, {Liu}, {Loots},
  {MacMahon}, {Malan}, {Malgas}, {Maree}, {Martinot}, {Mathison}, {Matsetela},
  {Mesinger}, {Morales}, {Neben}, {Patra}, {Pieterse}, {Pober}, {Razavi-Ghods},
  {Ringuette}, {Robnett}, {Rosie}, {Sell}, {Smith}, {Syce}, {Tegmark},
  {Thyagarajan}, {Williams}, \& {Zheng}}]{2017PASP..129d5001D}
{DeBoer}, D.~R., {Parsons}, A.~R., {Aguirre}, J.~E., {et~al.} 2017, \pasp, 129,
  045001, \dodoi{10.1088/1538-3873/129/974/045001}

\bibitem[{{Doughty} \& {Finlator}(2021)}]{2021MNRAS.505.2207D}
{Doughty}, C., \& {Finlator}, K. 2021, \mnras, 505, 2207,
  \dodoi{10.1093/mnras/stab1448}

\bibitem[{{Fan} {et~al.}(2006){Fan}, {Strauss}, {Becker}, {White}, {Gunn},
  {Knapp}, {Richards}, {Schneider}, {Brinkmann}, \&
  {Fukugita}}]{2006AJ....132..117F}
{Fan}, X., {Strauss}, M.~A., {Becker}, R.~H., {et~al.} 2006, \aj, 132, 117,
  \dodoi{10.1086/504836}

\bibitem[{{Finlator} {et~al.}(2011{\natexlab{a}}){Finlator}, {Dav{\'e}}, \&
  {{\"O}zel}}]{2011ApJ...743..169F}
{Finlator}, K., {Dav{\'e}}, R., \& {{\"O}zel}, F. 2011{\natexlab{a}}, \apj,
  743, 169, \dodoi{10.1088/0004-637X/743/2/169}

\bibitem[{{Finlator} {et~al.}(2011{\natexlab{b}}){Finlator}, {Dav{\'e}}, \&
  {{\"O}zel}}]{Finlator:2011}
---. 2011{\natexlab{b}}, \apj, 743, 169, \dodoi{10.1088/0004-637X/743/2/169}

\bibitem[{{Furlanetto} {et~al.}(2006){Furlanetto}, {McQuinn}, \&
  {Hernquist}}]{2006MNRAS.365..115F}
{Furlanetto}, S.~R., {McQuinn}, M., \& {Hernquist}, L. 2006, \mnras, 365, 115,
  \dodoi{10.1111/j.1365-2966.2005.09687.x}

\bibitem[{{Gardner} {et~al.}(2006){Gardner}, {Mather}, {Clampin}, {Doyon},
  {Greenhouse}, {Hammel}, {Hutchings}, {Jakobsen}, {Lilly}, {Long}, {Lunine},
  {McCaughrean}, {Mountain}, {Nella}, {Rieke}, {Rieke}, {Rix}, {Smith},
  {Sonneborn}, {Stiavelli}, {Stockman}, {Windhorst}, \&
  {Wright}}]{2006SSRv..123..485G}
{Gardner}, J.~P., {Mather}, J.~C., {Clampin}, M., {et~al.} 2006, \ssr, 123,
  485, \dodoi{10.1007/s11214-006-8315-7}

\bibitem[{{Genel} {et~al.}(2014){Genel}, {Vogelsberger}, {Springel}, {Sijacki},
  {Nelson}, {Snyder}, {Rodriguez-Gomez}, {Torrey}, \&
  {Hernquist}}]{2014MNRAS.445..175G}
{Genel}, S., {Vogelsberger}, M., {Springel}, V., {et~al.} 2014, \mnras, 445,
  175, \dodoi{10.1093/mnras/stu1654}

\bibitem[{{Gnedin}(2014)}]{2014ApJ...793...29G}
{Gnedin}, N.~Y. 2014, \apj, 793, 29, \dodoi{10.1088/0004-637X/793/1/29}

\bibitem[{{Gnedin} \& {Kaurov}(2014)}]{2014ApJ...793...30G}
{Gnedin}, N.~Y., \& {Kaurov}, A.~A. 2014, \apj, 793, 30,
  \dodoi{10.1088/0004-637X/793/1/30}

\bibitem[{{Haardt} \& {Madau}(2012)}]{Haardt:2012}
{Haardt}, F., \& {Madau}, P. 2012, \apj, 746, 125,
  \dodoi{10.1088/0004-637X/746/2/125}

\bibitem[{{Hassan} {et~al.}(2016){Hassan}, {Dav{\'e}}, {Finlator}, \&
  {Santos}}]{Hassan:2016}
{Hassan}, S., {Dav{\'e}}, R., {Finlator}, K., \& {Santos}, M.~G. 2016, \mnras,
  457, 1550, \dodoi{10.1093/mnras/stv3001}

\bibitem[{{Hopkins}(2015)}]{Hopkins:2015}
{Hopkins}, P.~F. 2015, \mnras, 450, 53, \dodoi{10.1093/mnras/stv195}

\bibitem[{{Hopkins}(2017)}]{Hopkins:2018}
---. 2017, ArXiv e-prints.
\newblock \doarXiv{1712.01294}

\bibitem[{{Iliev} {et~al.}(2014){Iliev}, {Mellema}, {Ahn}, {Shapiro}, {Mao}, \&
  {Pen}}]{2014MNRAS.439..725I}
{Iliev}, I.~T., {Mellema}, G., {Ahn}, K., {et~al.} 2014, \mnras, 439, 725,
  \dodoi{10.1093/mnras/stt2497}

\bibitem[{{Ivezi{\'c}} {et~al.}(2019){Ivezi{\'c}}, {Kahn}, {Tyson}, {Abel},
  {Acosta}, {Allsman}, {Alonso}, {AlSayyad}, {Anderson}, {Andrew}, {Angel},
  {Angeli}, {Ansari}, {Antilogus}, {Araujo}, {Armstrong}, {Arndt}, {Astier},
  {Aubourg}, {Auza}, {Axelrod}, {Bard}, {Barr}, {Barrau}, {Bartlett}, {Bauer},
  {Bauman}, {Baumont}, {Bechtol}, {Bechtol}, {Becker}, {Becla}, {Beldica},
  {Bellavia}, {Bianco}, {Biswas}, {Blanc}, {Blazek}, {Blandford}, {Bloom},
  {Bogart}, {Bond}, {Booth}, {Borgland}, {Borne}, {Bosch}, {Boutigny},
  {Brackett}, {Bradshaw}, {Brandt}, {Brown}, {Bullock}, {Burchat}, {Burke},
  {Cagnoli}, {Calabrese}, {Callahan}, {Callen}, {Carlin}, {Carlson},
  {Chandrasekharan}, {Charles-Emerson}, {Chesley}, {Cheu}, {Chiang}, {Chiang},
  {Chirino}, {Chow}, {Ciardi}, {Claver}, {Cohen-Tanugi}, {Cockrum}, {Coles},
  {Connolly}, {Cook}, {Cooray}, {Covey}, {Cribbs}, {Cui}, {Cutri}, {Daly},
  {Daniel}, {Daruich}, {Daubard}, {Daues}, {Dawson}, {Delgado}, {Dellapenna},
  {de Peyster}, {de Val-Borro}, {Digel}, {Doherty}, {Dubois},
  {Dubois-Felsmann}, {Durech}, {Economou}, {Eifler}, {Eracleous}, {Emmons},
  {Fausti Neto}, {Ferguson}, {Figueroa}, {Fisher-Levine}, {Focke}, {Foss},
  {Frank}, {Freemon}, {Gangler}, {Gawiser}, {Geary}, {Gee}, {Geha}, {Gessner},
  {Gibson}, {Gilmore}, {Glanzman}, {Glick}, {Goldina}, {Goldstein}, {Goodenow},
  {Graham}, {Gressler}, {Gris}, {Guy}, {Guyonnet}, {Haller}, {Harris},
  {Hascall}, {Haupt}, {Hernandez}, {Herrmann}, {Hileman}, {Hoblitt}, {Hodgson},
  {Hogan}, {Howard}, {Huang}, {Huffer}, {Ingraham}, {Innes}, {Jacoby}, {Jain},
  {Jammes}, {Jee}, {Jenness}, {Jernigan}, {Jevremovi{\'c}}, {Johns}, {Johnson},
  {Johnson}, {Jones}, {Juramy-Gilles}, {Juri{\'c}}, {Kalirai}, {Kallivayalil},
  {Kalmbach}, {Kantor}, {Karst}, {Kasliwal}, {Kelly}, {Kessler}, {Kinnison},
  {Kirkby}, {Knox}, {Kotov}, {Krabbendam}, {Krughoff}, {Kub{\'a}nek},
  {Kuczewski}, {Kulkarni}, {Ku}, {Kurita}, {Lage}, {Lambert}, {Lange},
  {Langton}, {Le Guillou}, {Levine}, {Liang}, {Lim}, {Lintott}, {Long},
  {Lopez}, {Lotz}, {Lupton}, {Lust}, {MacArthur}, {Mahabal}, {Mandelbaum},
  {Markiewicz}, {Marsh}, {Marshall}, {Marshall}, {May}, {McKercher}, {McQueen},
  {Meyers}, {Migliore}, {Miller}, {Mills}, {Miraval}, {Moeyens}, {Moolekamp},
  {Monet}, {Moniez}, {Monkewitz}, {Montgomery}, {Morrison}, {Mueller},
  {Muller}, {Mu{\~n}oz Arancibia}, {Neill}, {Newbry}, {Nief}, {Nomerotski},
  {Nordby}, {O'Connor}, {Oliver}, {Olivier}, {Olsen}, {O'Mullane}, {Ortiz},
  {Osier}, {Owen}, {Pain}, {Palecek}, {Parejko}, {Parsons}, {Pease},
  {Peterson}, {Peterson}, {Petravick}, {Libby Petrick}, {Petry},
  {Pierfederici}, {Pietrowicz}, {Pike}, {Pinto}, {Plante}, {Plate}, {Plutchak},
  {Price}, {Prouza}, {Radeka}, {Rajagopal}, {Rasmussen}, {Regnault}, {Reil},
  {Reiss}, {Reuter}, {Ridgway}, {Riot}, {Ritz}, {Robinson}, {Roby}, {Roodman},
  {Rosing}, {Roucelle}, {Rumore}, {Russo}, {Saha}, {Sassolas}, {Schalk},
  {Schellart}, {Schindler}, {Schmidt}, {Schneider}, {Schneider}, {Schoening},
  {Schumacher}, {Schwamb}, {Sebag}, {Selvy}, {Sembroski}, {Seppala}, {Serio},
  {Serrano}, {Shaw}, {Shipsey}, {Sick}, {Silvestri}, {Slater}, {Smith},
  {Smith}, {Sobhani}, {Soldahl}, {Storrie-Lombardi}, {Stover}, {Strauss},
  {Street}, {Stubbs}, {Sullivan}, {Sweeney}, {Swinbank}, {Szalay}, {Takacs},
  {Tether}, {Thaler}, {Thayer}, {Thomas}, {Thornton}, {Thukral}, {Tice},
  {Trilling}, {Turri}, {Van Berg}, {Vanden Berk}, {Vetter}, {Virieux},
  {Vucina}, {Wahl}, {Walkowicz}, {Walsh}, {Walter}, {Wang}, {Wang}, {Warner},
  {Wiecha}, {Willman}, {Winters}, {Wittman}, {Wolff}, {Wood-Vasey}, {Wu},
  {Xin}, {Yoachim}, \& {Zhan}}]{2019ApJ...873..111I}
{Ivezi{\'c}}, {\v{Z}}., {Kahn}, S.~M., {Tyson}, J.~A., {et~al.} 2019, \apj,
  873, 111, \dodoi{10.3847/1538-4357/ab042c}

\bibitem[{{Iyer} {et~al.}(2020){Iyer}, {Tacchella}, {Genel}, {Hayward},
  {Hernquist}, {Brooks}, {Caplar}, {Dav{\'e}}, {Diemer}, {Forbes}, {Gawiser},
  {Somerville}, \& {Starkenburg}}]{2020MNRAS.498..430I}
{Iyer}, K.~G., {Tacchella}, S., {Genel}, S., {et~al.} 2020, \mnras, 498, 430,
  \dodoi{10.1093/mnras/staa2150}

\bibitem[{{Kannan} {et~al.}(2022){Kannan}, {Garaldi}, {Smith}, {Pakmor},
  {Springel}, {Vogelsberger}, \& {Hernquist}}]{2022MNRAS.511.4005K}
{Kannan}, R., {Garaldi}, E., {Smith}, A., {et~al.} 2022, \mnras, 511, 4005,
  \dodoi{10.1093/mnras/stab3710}

\bibitem[{{Kannan} {et~al.}(2019){Kannan}, {Vogelsberger}, {Marinacci},
  {McKinnon}, {Pakmor}, \& {Springel}}]{2019MNRAS.485..117K}
{Kannan}, R., {Vogelsberger}, M., {Marinacci}, F., {et~al.} 2019, \mnras, 485,
  117, \dodoi{10.1093/mnras/stz287}

\bibitem[{{Katz} {et~al.}(2021){Katz}, {Martin-Alvarez}, {Rosdahl}, {Kimm},
  {Blaizot}, {Haehnelt}, {Michel-Dansac}, {Garel}, {O{\~n}orbe}, {Devriendt},
  {Slyz}, {Attia}, \& {Teyssier}}]{2021MNRAS.507.1254K}
{Katz}, H., {Martin-Alvarez}, S., {Rosdahl}, J., {et~al.} 2021, \mnras, 507,
  1254, \dodoi{10.1093/mnras/stab2148}

\bibitem[{{Kaviraj} {et~al.}(2017){Kaviraj}, {Laigle}, {Kimm}, {Devriendt},
  {Dubois}, {Pichon}, {Slyz}, {Chisari}, \& {Peirani}}]{2017MNRAS.467.4739K}
{Kaviraj}, S., {Laigle}, C., {Kimm}, T., {et~al.} 2017, \mnras, 467, 4739,
  \dodoi{10.1093/mnras/stx126}

\bibitem[{{Keating} {et~al.}(2020){Keating}, {Weinberger}, {Kulkarni},
  {Haehnelt}, {Chardin}, \& {Aubert}}]{2020MNRAS.491.1736K}
{Keating}, L.~C., {Weinberger}, L.~H., {Kulkarni}, G., {et~al.} 2020, \mnras,
  491, 1736, \dodoi{10.1093/mnras/stz3083}

\bibitem[{{Krumholz} {et~al.}(2009{\natexlab{a}}){Krumholz}, {McKee}, \&
  {Tumlinson}}]{2009ApJ...693..216K}
{Krumholz}, M.~R., {McKee}, C.~F., \& {Tumlinson}, J. 2009{\natexlab{a}}, \apj,
  693, 216, \dodoi{10.1088/0004-637X/693/1/216}

\bibitem[{{Krumholz} {et~al.}(2009{\natexlab{b}}){Krumholz}, {McKee}, \&
  {Tumlinson}}]{Krumholz:2009}
---. 2009{\natexlab{b}}, \apj, 693, 216, \dodoi{10.1088/0004-637X/693/1/216}

\bibitem[{{Kulkarni} {et~al.}(2019){Kulkarni}, {Keating}, {Haehnelt}, {Bosman},
  {Puchwein}, {Chardin}, \& {Aubert}}]{2019MNRAS.485L..24K}
{Kulkarni}, G., {Keating}, L.~C., {Haehnelt}, M.~G., {et~al.} 2019, \mnras,
  485, L24, \dodoi{10.1093/mnrasl/slz025}

\bibitem[{{Leroy} {et~al.}(2008){Leroy}, {Walter}, {Brinks}, {Bigiel}, {de
  Blok}, {Madore}, \& {Thornley}}]{Leroy:2008}
{Leroy}, A.~K., {Walter}, F., {Brinks}, E., {et~al.} 2008, \aj, 136, 2782,
  \dodoi{10.1088/0004-6256/136/6/2782}

\bibitem[{{Leung} {et~al.}(2020){Leung}, {Olsen}, {Somerville}, {Dav{\'e}},
  {Greve}, {Hayward}, {Narayanan}, \& {Popping}}]{2020ApJ...905..102L}
{Leung}, T.~K.~D., {Olsen}, K.~P., {Somerville}, R.~S., {et~al.} 2020, \apj,
  905, 102, \dodoi{10.3847/1538-4357/abc25e}

\bibitem[{{Li} {et~al.}(2019){Li}, {Narayanan}, \& {Dav{\'e}}}]{Li:2019}
{Li}, Q., {Narayanan}, D., \& {Dav{\'e}}, R. 2019, arXiv e-prints,
  arXiv:1906.09277.
\newblock \doarXiv{1906.09277}

\bibitem[{{Ma} {et~al.}(2020){Ma}, {Quataert}, {Wetzel}, {Hopkins},
  {Faucher-Gigu{\`e}re}, \& {Kere{\v{s}}}}]{2020MNRAS.498.2001M}
{Ma}, X., {Quataert}, E., {Wetzel}, A., {et~al.} 2020, \mnras, 498, 2001,
  \dodoi{10.1093/mnras/staa2404}

\bibitem[{{Majumdar} {et~al.}(2014){Majumdar}, {Mellema}, {Datta}, {Jensen},
  {Choudhury}, {Bharadwaj}, \& {Friedrich}}]{2014MNRAS.443.2843M}
{Majumdar}, S., {Mellema}, G., {Datta}, K.~K., {et~al.} 2014, \mnras, 443,
  2843, \dodoi{10.1093/mnras/stu1342}

\bibitem[{{McQuinn} {et~al.}(2007){McQuinn}, {Lidz}, {Zahn}, {Dutta},
  {Hernquist}, \& {Zaldarriaga}}]{2007MNRAS.377.1043M}
{McQuinn}, M., {Lidz}, A., {Zahn}, O., {et~al.} 2007, \mnras, 377, 1043,
  \dodoi{10.1111/j.1365-2966.2007.11489.x}

\bibitem[{{Mellema} {et~al.}(2013){Mellema}, {Koopmans}, {Abdalla}, {Bernardi},
  {Ciardi}, {Daiboo}, {de Bruyn}, {Datta}, {Falcke}, {Ferrara}, {Iliev},
  {Iocco}, {Jeli{\'c}}, {Jensen}, {Joseph}, {Labroupoulos}, {Meiksin},
  {Mesinger}, {Offringa}, {Pandey}, {Pritchard}, {Santos}, {Schwarz},
  {Semelin}, {Vedantham}, {Yatawatta}, \& {Zaroubi}}]{2013ExA....36..235M}
{Mellema}, G., {Koopmans}, L. V.~E., {Abdalla}, F.~A., {et~al.} 2013,
  Experimental Astronomy, 36, 235, \dodoi{10.1007/s10686-013-9334-5}

\bibitem[{{Mesinger} {et~al.}(2011){Mesinger}, {Furlanetto}, \&
  {Cen}}]{2011MNRAS.411..955M}
{Mesinger}, A., {Furlanetto}, S., \& {Cen}, R. 2011, \mnras, 411, 955,
  \dodoi{10.1111/j.1365-2966.2010.17731.x}

\bibitem[{{Molaro} {et~al.}(2019){Molaro}, {Dav{\'e}}, {Hassan}, {Santos}, \&
  {Finlator}}]{2019MNRAS.489.5594M}
{Molaro}, M., {Dav{\'e}}, R., {Hassan}, S., {Santos}, M.~G., \& {Finlator}, K.
  2019, \mnras, 489, 5594, \dodoi{10.1093/mnras/stz2171}

\bibitem[{{Muratov} {et~al.}(2015){Muratov}, {Kere{\v s}},
  {Faucher-Gigu{\`e}re}, {Hopkins}, {Quataert}, \& {Murray}}]{Muratov:2015}
{Muratov}, A.~L., {Kere{\v s}}, D., {Faucher-Gigu{\`e}re}, C.-A., {et~al.}
  2015, \mnras, 454, 2691, \dodoi{10.1093/mnras/stv2126}

\bibitem[{{Ocvirk} {et~al.}(2021){Ocvirk}, {Lewis}, {Gillet}, {Chardin},
  {Aubert}, {Deparis}, \& {Thelie}}]{2021arXiv210501663O}
{Ocvirk}, P., {Lewis}, J. S.~W., {Gillet}, N., {et~al.} 2021, arXiv e-prints,
  arXiv:2105.01663.
\newblock \doarXiv{2105.01663}

\bibitem[{{Ocvirk} {et~al.}(2020){Ocvirk}, {Aubert}, {Sorce}, {Shapiro},
  {Deparis}, {Dawoodbhoy}, {Lewis}, {Teyssier}, {Yepes}, {Gottl{\"o}ber},
  {Ahn}, {Iliev}, \& {Hoffman}}]{2020MNRAS.496.4087O}
{Ocvirk}, P., {Aubert}, D., {Sorce}, J.~G., {et~al.} 2020, \mnras, 496, 4087,
  \dodoi{10.1093/mnras/staa1266}

\bibitem[{{Pillepich} {et~al.}(2018){Pillepich}, {Springel}, {Nelson}, {Genel},
  {Naiman}, {Pakmor}, {Hernquist}, {Torrey}, {Vogelsberger}, {Weinberger}, \&
  {Marinacci}}]{2018MNRAS.473.4077P}
{Pillepich}, A., {Springel}, V., {Nelson}, D., {et~al.} 2018, \mnras, 473,
  4077, \dodoi{10.1093/mnras/stx2656}

\bibitem[{{Planck Collaboration} {et~al.}(2016){Planck Collaboration}, {Ade},
  {Aghanim}, {Arnaud}, {Ashdown}, {Aumont}, {Baccigalupi}, {Banday},
  {Barreiro}, {Bartlett}, {Bartolo}, {Battaner}, {Battye}, {Benabed},
  {Beno{\^\i}t}, {Benoit-L{\'e}vy}, {Bernard}, {Bersanelli}, {Bielewicz},
  {Bock}, {Bonaldi}, {Bonavera}, {Bond}, {Borrill}, {Bouchet}, {Boulanger},
  {Bucher}, {Burigana}, {Butler}, {Calabrese}, {Cardoso}, {Catalano},
  {Challinor}, {Chamballu}, {Chary}, {Chiang}, {Chluba}, {Christensen},
  {Church}, {Clements}, {Colombi}, {Colombo}, {Combet}, {Coulais}, {Crill},
  {Curto}, {Cuttaia}, {Danese}, {Davies}, {Davis}, {de Bernardis}, {de Rosa},
  {de Zotti}, {Delabrouille}, {D{\'e}sert}, {Di Valentino}, {Dickinson},
  {Diego}, {Dolag}, {Dole}, {Donzelli}, {Dor{\'e}}, {Douspis}, {Ducout},
  {Dunkley}, {Dupac}, {Efstathiou}, {Elsner}, {En{\ss}lin}, {Eriksen},
  {Farhang}, {Fergusson}, {Finelli}, {Forni}, {Frailis}, {Fraisse},
  {Franceschi}, {Frejsel}, {Galeotta}, {Galli}, {Ganga}, {Gauthier}, {Gerbino},
  {Ghosh}, {Giard}, {Giraud-H{\'e}raud}, {Giusarma}, {Gjerl{\o}w},
  {Gonz{\'a}lez-Nuevo}, {G{\'o}rski}, {Gratton}, {Gregorio}, {Gruppuso},
  {Gudmundsson}, {Hamann}, {Hansen}, {Hanson}, {Harrison}, {Helou},
  {Henrot-Versill{\'e}}, {Hern{\'a}ndez-Monteagudo}, {Herranz}, {Hildebrandt},
  {Hivon}, {Hobson}, {Holmes}, {Hornstrup}, {Hovest}, {Huang}, {Huffenberger},
  {Hurier}, {Jaffe}, {Jaffe}, {Jones}, {Juvela}, {Keih{\"a}nen}, {Keskitalo},
  {Kisner}, {Kneissl}, {Knoche}, {Knox}, {Kunz}, {Kurki-Suonio}, {Lagache},
  {L{\"a}hteenm{\"a}ki}, {Lamarre}, {Lasenby}, {Lattanzi}, {Lawrence}, {Leahy},
  {Leonardi}, {Lesgourgues}, {Levrier}, {Lewis}, {Liguori}, {Lilje},
  {Linden-V{\o}rnle}, {L{\'o}pez-Caniego}, {Lubin}, {Mac{\'\i}as-P{\'e}rez},
  {Maggio}, {Maino}, {Mandolesi}, {Mangilli}, {Marchini}, {Maris}, {Martin},
  {Martinelli}, {Mart{\'\i}nez-Gonz{\'a}lez}, {Masi}, {Matarrese}, {McGehee},
  {Meinhold}, {Melchiorri}, {Melin}, {Mendes}, {Mennella}, {Migliaccio},
  {Millea}, {Mitra}, {Miville-Desch{\^e}nes}, {Moneti}, {Montier}, {Morgante},
  {Mortlock}, {Moss}, {Munshi}, {Murphy}, {Naselsky}, {Nati}, {Natoli},
  {Netterfield}, {N{\o}rgaard-Nielsen}, {Noviello}, {Novikov}, {Novikov},
  {Oxborrow}, {Paci}, {Pagano}, {Pajot}, {Paladini}, {Paoletti}, {Partridge},
  {Pasian}, {Patanchon}, {Pearson}, {Perdereau}, {Perotto}, {Perrotta},
  {Pettorino}, {Piacentini}, {Piat}, {Pierpaoli}, {Pietrobon}, {Plaszczynski},
  {Pointecouteau}, {Polenta}, {Popa}, {Pratt}, {Pr{\'e}zeau}, {Prunet},
  {Puget}, {Rachen}, {Reach}, {Rebolo}, {Reinecke}, {Remazeilles}, {Renault},
  {Renzi}, {Ristorcelli}, {Rocha}, {Rosset}, {Rossetti}, {Roudier},
  {Rouill{\'e} d'Orfeuil}, {Rowan-Robinson}, {Rubi{\~n}o-Mart{\'\i}n},
  {Rusholme}, {Said}, {Salvatelli}, {Salvati}, {Sandri}, {Santos},
  {Savelainen}, {Savini}, {Scott}, {Seiffert}, {Serra}, {Shellard}, {Spencer},
  {Spinelli}, {Stolyarov}, {Stompor}, {Sudiwala}, {Sunyaev}, {Sutton},
  {Suur-Uski}, {Sygnet}, {Tauber}, {Terenzi}, {Toffolatti}, {Tomasi},
  {Tristram}, {Trombetti}, {Tucci}, {Tuovinen}, {T{\"u}rler}, {Umana},
  {Valenziano}, {Valiviita}, {Van Tent}, {Vielva}, {Villa}, {Wade}, {Wandelt},
  {Wehus}, {White}, {White}, {Wilkinson}, {Yvon}, {Zacchei}, \&
  {Zonca}}]{2016A&A...594A..13P}
{Planck Collaboration}, {Ade}, P.~A.~R., {Aghanim}, N., {et~al.} 2016, \aap,
  594, A13, \dodoi{10.1051/0004-6361/201525830}

\bibitem[{{Press} \& {Schechter}(1974)}]{1974ApJ...187..425P}
{Press}, W.~H., \& {Schechter}, P. 1974, \apj, 187, 425, \dodoi{10.1086/152650}

\bibitem[{{Rahmati} {et~al.}(2013){Rahmati}, {Pawlik}, Raicevic, \&
  {Schaye}}]{Rahmati:2013}
{Rahmati}, A., {Pawlik}, A.~H., Raicevic, M., \& {Schaye}, J. 2013, \mnras,
  430, 2427, \dodoi{10.1093/mnras/stt066}

\bibitem[{{Rosdahl} {et~al.}(2018){Rosdahl}, {Katz}, {Blaizot}, {Kimm},
  {Michel-Dansac}, {Garel}, {Haehnelt}, {Ocvirk}, \&
  {Teyssier}}]{2018MNRAS.479..994R}
{Rosdahl}, J., {Katz}, H., {Blaizot}, J., {et~al.} 2018, \mnras, 479, 994,
  \dodoi{10.1093/mnras/sty1655}

\bibitem[{{Santos} {et~al.}(2010){Santos}, {Ferramacho}, {Silva}, {Amblard}, \&
  {Cooray}}]{2010MNRAS.406.2421S}
{Santos}, M.~G., {Ferramacho}, L., {Silva}, M.~B., {Amblard}, A., \& {Cooray},
  A. 2010, \mnras, 406, 2421, \dodoi{10.1111/j.1365-2966.2010.16898.x}

\bibitem[{{Schaerer}(2003)}]{2003A&A...397..527S}
{Schaerer}, D. 2003, \aap, 397, 527, \dodoi{10.1051/0004-6361:20021525}

\bibitem[{{Schaye} {et~al.}(2015){Schaye}, {Crain}, {Bower}, {Furlong},
  {Schaller}, {Theuns}, {Dalla Vecchia}, {Frenk}, {McCarthy}, {Helly},
  {Jenkins}, {Rosas-Guevara}, {White}, {Baes}, {Booth}, {Camps}, {Navarro},
  {Qu}, {Rahmati}, {Sawala}, {Thomas}, \& {Trayford}}]{2015MNRAS.446..521S}
{Schaye}, J., {Crain}, R.~A., {Bower}, R.~G., {et~al.} 2015, \mnras, 446, 521,
  \dodoi{10.1093/mnras/stu2058}

\bibitem[{{Smith} {et~al.}(2017){Smith}, {Bryan}, {Glover}, {Goldbaum}, {Turk},
  {Regan}, {Wise}, {Schive}, {Abel}, {Emerick}, {O'Shea}, {Anninos}, {Hummels},
  \& {Khochfar}}]{Smith:2017}
{Smith}, B.~D., {Bryan}, G.~L., {Glover}, S.~C.~O., {et~al.} 2017, \mnras, 466,
  2217, \dodoi{10.1093/mnras/stw3291}

\bibitem[{{Sobacchi} \& {Mesinger}(2014)}]{2014MNRAS.440.1662S}
{Sobacchi}, E., \& {Mesinger}, A. 2014, \mnras, 440, 1662,
  \dodoi{10.1093/mnras/stu377}

\bibitem[{{Spergel} {et~al.}(2015){Spergel}, {Gehrels}, {Baltay}, {Bennett},
  {Breckinridge}, {Donahue}, {Dressler}, {Gaudi}, {Greene}, {Guyon}, {Hirata},
  {Kalirai}, {Kasdin}, {Macintosh}, {Moos}, {Perlmutter}, {Postman},
  {Rauscher}, {Rhodes}, {Wang}, {Weinberg}, {Benford}, {Hudson}, {Jeong},
  {Mellier}, {Traub}, {Yamada}, {Capak}, {Colbert}, {Masters}, {Penny},
  {Savransky}, {Stern}, {Zimmerman}, {Barry}, {Bartusek}, {Carpenter}, {Cheng},
  {Content}, {Dekens}, {Demers}, {Grady}, {Jackson}, {Kuan}, {Kruk}, {Melton},
  {Nemati}, {Parvin}, {Poberezhskiy}, {Peddie}, {Ruffa}, {Wallace}, {Whipple},
  {Wollack}, \& {Zhao}}]{2015arXiv150303757S}
{Spergel}, D., {Gehrels}, N., {Baltay}, C., {et~al.} 2015, arXiv e-prints,
  arXiv:1503.03757.
\newblock \doarXiv{1503.03757}

\bibitem[{{Springel} \& {Hernquist}(2003)}]{2003MNRAS.339..289S}
{Springel}, V., \& {Hernquist}, L. 2003, \mnras, 339, 289,
  \dodoi{10.1046/j.1365-8711.2003.06206.x}

\bibitem[{{Stark}(2016)}]{2016ARA&A..54..761S}
{Stark}, D.~P. 2016, \araa, 54, 761,
  \dodoi{10.1146/annurev-astro-081915-023417}

\bibitem[{{van Haarlem} {et~al.}(2013){van Haarlem}, {Wise}, {Gunst}, {Heald},
  {McKean}, {Hessels}, {de Bruyn}, {Nijboer}, {Swinbank}, {Fallows},
  {Brentjens}, {Nelles}, {Beck}, {Falcke}, {Fender}, {H{\"o}randel},
  {Koopmans}, {Mann}, {Miley}, {R{\"o}ttgering}, {Stappers}, {Wijers},
  {Zaroubi}, {van den Akker}, {Alexov}, {Anderson}, {Anderson}, {van Ardenne},
  {Arts}, {Asgekar}, {Avruch}, {Batejat}, {B{\"a}hren}, {Bell}, {Bell}, {van
  Bemmel}, {Bennema}, {Bentum}, {Bernardi}, {Best}, {B{\^\i}rzan}, {Bonafede},
  {Boonstra}, {Braun}, {Bregman}, {Breitling}, {van de Brink}, {Broderick},
  {Broekema}, {Brouw}, {Br{\"u}ggen}, {Butcher}, {van Cappellen}, {Ciardi},
  {Coenen}, {Conway}, {Coolen}, {Corstanje}, {Damstra}, {Davies}, {Deller},
  {Dettmar}, {van Diepen}, {Dijkstra}, {Donker}, {Doorduin}, {Dromer}, {Drost},
  {van Duin}, {Eisl{\"o}ffel}, {van Enst}, {Ferrari}, {Frieswijk}, {Gankema},
  {Garrett}, {de Gasperin}, {Gerbers}, {de Geus}, {Grie{\ss}meier}, {Grit},
  {Gruppen}, {Hamaker}, {Hassall}, {Hoeft}, {Holties}, {Horneffer}, {van der
  Horst}, {van Houwelingen}, {Huijgen}, {Iacobelli}, {Intema}, {Jackson},
  {Jelic}, {de Jong}, {Juette}, {Kant}, {Karastergiou}, {Koers}, {Kollen},
  {Kondratiev}, {Kooistra}, {Koopman}, {Koster}, {Kuniyoshi}, {Kramer},
  {Kuper}, {Lambropoulos}, {Law}, {van Leeuwen}, {Lemaitre}, {Loose}, {Maat},
  {Macario}, {Markoff}, {Masters}, {McFadden}, {McKay-Bukowski}, {Meijering},
  {Meulman}, {Mevius}, {Middelberg}, {Millenaar}, {Miller-Jones}, {Mohan},
  {Mol}, {Morawietz}, {Morganti}, {Mulcahy}, {Mulder}, {Munk}, {Nieuwenhuis},
  {van Nieuwpoort}, {Noordam}, {Norden}, {Noutsos}, {Offringa}, {Olofsson},
  {Omar}, {Orr{\'u}}, {Overeem}, {Paas}, {Pandey-Pommier}, {Pandey}, {Pizzo},
  {Polatidis}, {Rafferty}, {Rawlings}, {Reich}, {de Reijer}, {Reitsma},
  {Renting}, {Riemers}, {Rol}, {Romein}, {Roosjen}, {Ruiter}, {Scaife}, {van
  der Schaaf}, {Scheers}, {Schellart}, {Schoenmakers}, {Schoonderbeek},
  {Serylak}, {Shulevski}, {Sluman}, {Smirnov}, {Sobey}, {Spreeuw}, {Steinmetz},
  {Sterks}, {Stiepel}, {Stuurwold}, {Tagger}, {Tang}, {Tasse}, {Thomas},
  {Thoudam}, {Toribio}, {van der Tol}, {Usov}, {van Veelen}, {van der Veen},
  {ter Veen}, {Verbiest}, {Vermeulen}, {Vermaas}, {Vocks}, {Vogt}, {de Vos},
  {van der Wal}, {van Weeren}, {Weggemans}, {Weltevrede}, {White}, {Wijnholds},
  {Wilhelmsson}, {Wucknitz}, {Yatawatta}, {Zarka}, {Zensus}, \& {van
  Zwieten}}]{2013A&A...556A...2V}
{van Haarlem}, M.~P., {Wise}, M.~W., {Gunst}, A.~W., {et~al.} 2013, \aap, 556,
  A2, \dodoi{10.1051/0004-6361/201220873}

\bibitem[{{Whitler} {et~al.}(2020){Whitler}, {Mason}, {Ren}, {Dijkstra},
  {Mesinger}, {Pentericci}, {Trenti}, \& {Treu}}]{2020MNRAS.495.3602W}
{Whitler}, L.~R., {Mason}, C.~A., {Ren}, K., {et~al.} 2020, \mnras, 495, 3602,
  \dodoi{10.1093/mnras/staa1178}

\bibitem[{{Wu} {et~al.}(2020){Wu}, {Dav{\'e}}, {Tacchella}, \&
  {Lotz}}]{2020MNRAS.494.5636W}
{Wu}, X., {Dav{\'e}}, R., {Tacchella}, S., \& {Lotz}, J. 2020, \mnras, 494,
  5636, \dodoi{10.1093/mnras/staa1044}

\bibitem[{{Wu} {et~al.}(2021){Wu}, {McQuinn}, \&
  {Eisenstein}}]{2021JCAP...02..042W}
{Wu}, X., {McQuinn}, M., \& {Eisenstein}, D. 2021, \jcap, 2021, 042,
  \dodoi{10.1088/1475-7516/2021/02/042}

\bibitem[{{Zahn} {et~al.}(2011){Zahn}, {Mesinger}, {McQuinn}, {Trac}, {Cen}, \&
  {Hernquist}}]{2011MNRAS.414..727Z}
{Zahn}, O., {Mesinger}, A., {McQuinn}, M., {et~al.} 2011, \mnras, 414, 727,
  \dodoi{10.1111/j.1365-2966.2011.18439.x}

\end{thebibliography}
\bibliographystyle{aasjournal}

\end{document}